\documentclass[traditabstract]{aa}

\usepackage{txfonts}
\usepackage{graphicx}
\usepackage{epsfig,graphics,lscape,url}
\usepackage{natbib}
\bibpunct{(}{)}{;}{a}{}{,}
\usepackage{longtable}

\begin{document}

\title{Atomic data for neutron-capture elements}
\subtitle{III.  Charge transfer rate coefficients for low-charge ions of Ge, Se, Br, Kr, Rb, and Xe}

\author{N.\ C.\ Sterling$^{\ref{inst1}}$\thanks{NSF Astronomy and Astrophysics Postdoctoral Fellow} \and P.\ C.\ Stancil$^{\ref{inst2}}$}

\institute{Michigan State University, Department of Physics and Astronomy, 3248 Biomedical Physical Sciences, East Lansing, MI, 48824-2320, USA, \email{sterling@pa.msu.edu} \label{inst1}
\and
Department of Physics and Astronomy and the Center for Simulational Physics, University of Georgia, Athens, GA 30602-2451, USA, \email{stancil@physast.uga.edu}\label{inst2}}


\abstract{We present total and final-state resolved charge transfer (CT) rate coefficients for low-charge Ge, Se, Br, Kr, Rb, and Xe ions reacting with neutral hydrogen over the temperature range 10$^2$--10$^6$~K.  Each of these elements has been detected in ionized astrophysical nebulae, particularly planetary nebulae.  CT rate coefficients are a key ingredient for the ionization equilibrium solutions needed to determine total elemental abundances from those of the observed ions.  A multi-channel Landau Zener approach was used to compute rate coefficients for projectile ions with charges $q=2$--5, and for singly-charged ions the Demkov approximation was utilized.  Our results for five-times ionized species are lower limits, due to the incompleteness of level energies in the NIST database.  In addition, we computed rate coefficients for charge transfer ionization reactions between the neutral species of the above six elements and ionized hydrogen.  The resulting total and state-resolved CT rate coefficients are tabulated and available at the CDS.  In tandem with our concurrent investigations of other important atomic processes in photoionized nebulae, this work will enable robust investigations of neutron-capture element abundances and nucleosynthesis via nebular spectroscopy.}


\maketitle

\titlerunning{Atomic Data for Neutron-Capture Elements III}
\authorrunning{N.\ C.\ Sterling \& P.\ C.\ Stancil}

\section{Introduction} \label{intro}

The detection of \emph{n}-capture elements (atomic number $Z>30$) in ionized astrophysical nebulae has underscored the need for atomic data required to accurately solve for the ionization equilibrium of these species.  Trans-iron element emission lines have been detected in objects ranging from planetary nebulae \citep[e.g.,][]{zhang05, sharpee07, sterling07, sterling08, sterling09, otsuka10b, otsuka11} to H~II regions \citep[e.g.,][and references therein]{baldwin00, blum08}, and the interstellar medium of other galaxies \citep{vanzi08}.  Determinations of their abundances have important implications for the production and chemical evolution of \emph{n}-capture nuclides in the Universe \citep[e.g.,][]{busso99, sneden08, sterling08}.  Se, Kr, and Xe are the most widely detected \emph{n}-capture elements in ionized nebulae \citep[e.g,][]{dinerstein01, sharpee07, sterling08, sterling09}, but other trans-iron elements have also been detected in nebular spectra, including Ge \citep{sterling02, sterling03}, Br, and Rb \citep{pb94, zhang05, sharpee07, sterling09}.

Due to the limited number of detectable \emph{n}-capture element ions in individual nebulae, it is necessary to convert abundances of the observed ions to elemental abundances by correcting for unobserved ionization stages.  Numerical simulations of ionized nebulae present a robust and accurate method of deriving such ionization corrections, provided that accurate data for atomic processes affecting the ionization equilibrium of the modeled elements are available.  Since \emph{n}-capture element emission lines have been detected primarily in photoionized nebulae, the most important atomic data required for ionization balance determinations are photoionization cross sections and rate coefficients for radiative recombination, dielectronic recombination, and charge transfer (CT).  Unfortunately, these data are poorly if at all known for ions of trans-iron elements.

This paper is the third in a series presenting theoretical atomic data determinations relevant to nebular ionization balance solutions for low-charge \emph{n}-capture element ions.  In the first two \citep{sterling11b, sterling11c}, we furnished multi-configuration Breit-Pauli distorted-wave photoionization cross sections and radiative and dielectronic recombination rate coefficients for Se and Kr ions.

To address the need for CT data for low-charge ions of \emph{n}-capture elements, we present calculations based on the Landau-Zener (LZ) and Demkov approximations.  We consider reactions of the form 
\begin{equation}
\mathrm{X}^{q+} + \mathrm{H} \rightleftharpoons \mathrm{X}^{(q-1)+} + \mathrm{H}^+ + \Delta \mathrm{E}, 
\label{rxn}
\end{equation}
with $q=1$--5 and X~=~Ge, Se, Br, Kr, Rb, and Xe.  The forward reaction, CT recombination (in terms of the projectile X$^{q+}$), is exothermic for these elements when $q\geq 2$, and the reverse reaction for multiply-charged systems is generally unimportant at the temperatures of photoionized nebulae (near 10$^4$~K).  On the other hand, of these elements only Kr has a higher ionization potential than H, and hence CT recombination is endoergic for singly-ionized Ge, Se, Br, Rb, and Xe.  We therefore provide rate coefficients for both CT recombination of singly-charged ions and CT ionization of neutral species.

Our LZ calculations are multi-channel, in the sense that several exit channels for the X$^{(q-1)+}$ system are simultaneously considered.  This avoids the problem of overestimating cross sections when adding two-channel cross sections of the same molecular symmetry \citep{bd80, kingdon95}.  For singly-ionized atoms, CT usually cannot be treated in the LZ approximation, and hence we utilize the Demkov approximation to determine rate coefficients for those systems.

LZ calculations are generally accurate to within a factor of three for systems (and final channels) with large rate coefficients \citep[$\gtrsim 10^{-9}$~cm$^3$\,s$^{-1}$;][]{bd80, kf96}, but can be less accurate when the rate coefficient is small (though in that case, CT often is much less important than radiative and dielectronic recombination).  The quantum mechanical molecular-orbital close-coupling (QMOCC) approach \citep[e.g.,][]{kimura89, zygelman92, wang04} is the most accurate theoretical method for determining CT rate coefficients at the temperatures of interest.  However, given the time- and computationally-intensive nature of such calculations, we have chosen to first render LZ calculations for these systems.  We will test the sensitivity of abundance determinations to uncertainties in the CT rate coefficients (assumed to be roughly a factor of three) through numerical simulations of ionized nebulae.  Those systems which are most critical for ionization equilibrium solutions are targets for further investigation with QMOCC methods.

The structure of this paper is as follows.  In Sect.~\ref{calcs}, we discuss the details and methodology of our calculations.  We present the resulting total and final-state resolved CT rate coefficients in Sect.~\ref{results}, and summarize our investigation in Sect.~\ref{summ}.

\section{Calculations and methodology}\label{calcs}

\subsection{Multi-channel Landau Zener calculations}\label{lz}

For projectiles with charge $q=2$--5, we compute CT rate coefficients in the multi-channel Landau-Zener (MCLZ) approximation, based on the formalism of \citet{bd80} and \citet{janev83} with the low-energy modifications of \citet{chang79}.

While CT is a quasi-molecular problem, the basic details of the interaction can be understood from consideration of the initial channel (in which both the projectile X$^{q+}$ and target H are assumed to be in their ground states), the final channels (which can include ground and excited states of X$^{(q-1)+}$), and the interaction between them.  As the atoms approach each other, if the ionization potential of X$^{(q-1)+}$ exceeds that of H (as is the case for $q=2$--5 for all elements we consider), CT is exoergic and the diabatic potential curves can cross at an internuclear distance $R_{\mathrm{x}}$.  The CT reaction is usually driven by a radial coupling that connects initial and final channels corresponding to the same molecular symmetry of XH$^{q+}$.  Adiabatic potential curves of molecular states with the same symmetry cannot cross, and $R_{\mathrm{x}}$ corresponds to the radius at which the adiabatic potentials most closely approach (the ``avoided crossing'' distance).

In the LZ approximation, charge exchange is assumed to be localized at the internuclear distance $R_{\mathrm{x}}$, with the probability of exchange given by $2p(1-p)$, where $p=e^{-w}$ and
\begin{equation}
w = \frac{\pi ^2 [\Delta U(R_{\mathrm{x}})]^2}{hv\left[ \frac{d}{dr} (H_{11} - H_{22}) \right]_{R_{\mathrm{x}}}}.
\label{wdef}
\end{equation}
In this equation, $v$ is the relative radial velocity of the projectile and target, $H_{11}$ and $H_{22}$ are the incoming and outgoing diabatic potentials, and $\Delta U(R_{\rm x})$ is the separation of the adiabatic potentials at $R_{\mathrm{x}}$.  The cross section at energy $E$ is 
\begin{equation}
\sigma(E) = 4\pi R_{\mathrm{x}}^2p_0(1+\lambda ) \int_{1}^{\infty}(1-e^{-wx})e^{-wx}x^{-3}dx,
\label{xsec}
\end{equation}
where $\lambda=[H_{11}(\infty) - H_{11}(R_{\mathrm{x}})]/E_i$ \citep{bd80}.  The approach probability $p_0$ is determined from the orbital and spin angular momenta of the initial and final states, which can form $g_{if} = (2L_i+1)(2S_i+1)(2L_f+1)(2S_f+1)$ molecular states, and the spin $S$ and orbital angular momentum projection quantum number $\Lambda '$ of the molecular state \citep{herzberg50, dalgarno90}:
\begin{equation}
p_0 = (2S+1)(2-\delta_{0,\Lambda '})/g_{if}.
\end{equation}

The above describes a two-channel LZ interaction.  In MCLZ calculations, several capture channels are considered simultaneously, each with a unique value of $R_{\mathrm{x}}$, as the projectile approaches the target.  In this case, for $N$ final channels the probability for capture into a state $n$ is given by \citep{janev83}
\begin{eqnarray}
P_n &=& p_1p_2\dots p_n(1-p_n)[1+(p_{n+1}p_{n+2}\dots p_N)^2+ \nonumber \\
 & & (p_{n+1}p_{n+2}\dots p_{N-1})^2(1-p_N)^2+\dots \nonumber \\
 & & +p_{n+1}^2(1-p_{n+2})^2+(1-p_{n+1})^2].
\label{mclzprob}
\end{eqnarray}

LZ calculations require estimates of the incoming and outgoing potentials, the avoided crossing distance $R_{\mathrm{x}}$, and the energy separation of the adiabatic potentials at $R_{\mathrm{x}}$.  We use approximations for these quantities from \citet{bd80} as follows.  In atomic units,\footnote{All quantities in this paper are in atomic units unless otherwise noted.} the incoming potential is
\begin{equation}
H_{11}(R) = \frac{-\alpha_{\mathrm{H}}q^2}{2R^4} + Aqe^{-(0.8+\zeta )R},
\label{h11}
\end{equation}
where $\alpha_{\mathrm{H}}$ is the polarizability of H, $q$ the charge of the projectile, $R$ the internuclear distance, and $\zeta$ is the exponent of a single orbital wavefunction ($-1.0$ for H targets).  Typically set to 25, $A$ was reduced in some cases to yield physically reasonable behavior in the potential, which occasionally was spurious in the repulsive core (this generally only occurs when $R_{\mathrm{x}} \lesssim 4.5$--5~$a_0$, in which case the rate coefficient is usually small at photoionized plasma temperatures).  The outgoing potential is given by
\begin{equation}
H_{22}(R) = \frac{(q-1)}{R} - \Delta E,
\label{h22}
\end{equation}
where $\Delta E$ = IP(X$^{q+}$) -- IP(H) -- $E_{\mathrm{exc}}$ is the energy defect, IP is the ionization potential of the indicated system, and $E_{\mathrm{exc}}$ is the excitation energy of the final X$^{(q-1)+}$ state.

\begin{figure}
  \resizebox{\hsize}{!}{\includegraphics{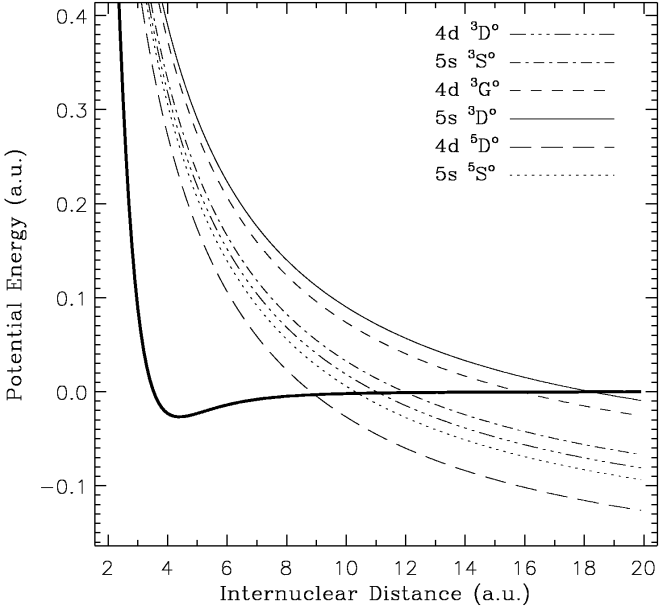}}
  \caption{Incoming diabatic potential $H_{11}$ for Kr$^{3+}$~+~H (thick solid line), and diabatic potentials $H_{22}$ for six exit channels of Kr$^{2+}$~+~H$^+$~+~$\Delta E$ corresponding to $^3 \Sigma^-$ and $^5 \Sigma^-$ molecular symmetries (all listed states have an [Ar]\,4s$^2$\,4p$^3$ core).  The intersections of the incoming and outgoing potentials correspond to the avoided crossing distances $R_{\mathrm{x}}$ for each exit channel.  The potential energies are shifted by taking $H_{11}(\infty)=0$.  } \label{kr3+_pots}
\end{figure}

The avoided crossing distances $R_{\mathrm{x}}$ are determined by equating Equations~\ref{h11} and \ref{h22}.  In Fig.~\ref{kr3+_pots}, we illustrate the diabatic potentials and $R_{\mathrm{x}}$ values for the final channels of the case X$^{q+}$~=~Kr$^{3+}$.  Note that we ignore fine structure in our calculations.


The final piece of necessary information, $\Delta U(R_{\mathrm{x}})$, is perhaps the greatest uncertainty in LZ calculations, as the formula provided by \cite{bd80} is an empirical fit to a small number of quantal calculations.  The value of $\Delta U(R_{\mathrm{x}})$ depends on whether the electron capture process is Type~I (the H $1s$ electron is transferred without affecting the core electrons of the projectile) or Type~II (electron capture is accompanied by the excitation of a core electron).  The formulae of \citet{bd80} for these two scenarios are:
\begin{eqnarray}
\Delta U_{\mathrm{I}} &=& 27.21R_{\mathrm{x}}^2e^{-\beta R_{\mathrm{x}}} \,\,\,\, \\
\Delta U_{\mathrm{II}} &=& 0.5e^{-0.4R_{\mathrm{x}}} \Delta U_{\mathrm{I}},
\end{eqnarray}
where $\beta=2\mathrm{IP(H)}=1.0$ in atomic units.  The potential for Type~II reactions is reduced relative to their Type~I counterparts due to the rearrangement of the core electrons.

\begin{figure}
  \resizebox{\hsize}{!}{\includegraphics{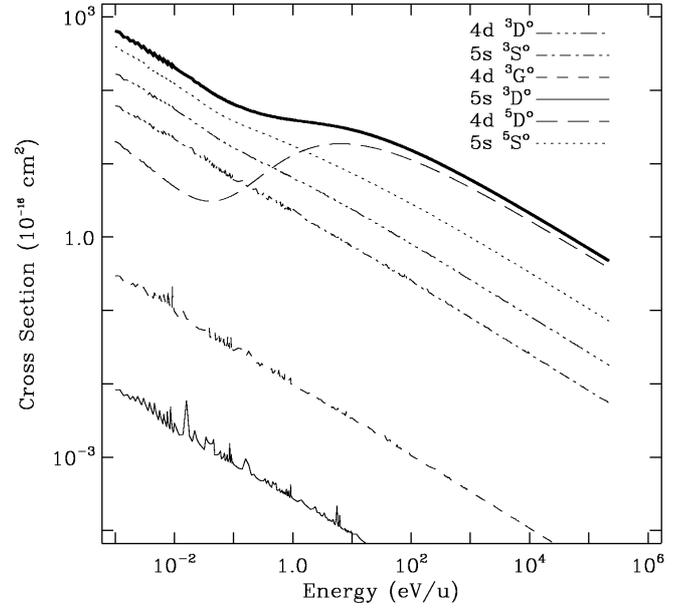}}
  \caption{Total (thick solid curve) and final-state resolved cross sections for CT recombination of Kr$^{3+}$.  Energies are given in eV/u, where u is the atomic mass unit.} \label{kr3+_xsec}
\end{figure}

Once the quantities $H_{11}$, $H_{22}$, $R_{\mathrm{x}}$, and $\Delta U(R_{\mathrm{x}})$ are determined, then the cross section for each final channel can be determined from Equation~\ref{xsec}, as illustrated for X$^{q+}=$~Kr$^{3+}$ in Fig.~\ref{kr3+_xsec}.  These are integrated over a Maxwellian velocity distribution to derive final-state resolved rate coefficients.

\subsection{Demkov calculations}\label{demkov}

In the case of singly-charged ions ($q=1$ in Equation~\ref{rxn}), the LZ approximation cannot be applied since the final channels do not have Coulomb potentials (hence there are no avoided crossings).  The reaction proceeds through radial coupling connecting molecular states with the same symmetry as in the LZ calculations, but the coupling is not localized and can be represented with an exponential form.  This type of interaction can be described by the Demkov approximation \citep{demkov64, olson72, swartz94}.

The reaction is assumed to occur at an interaction distance that we label $R_{\rm x}$ due to its analogous nature to the avoided crossing distance in LZ calculations.  $R_{\rm x}$ is given by equating $\Delta U(R)$ with \citep{swartz94}
\begin{equation}
\Delta H(R) = H_{11}(\mathrm{X}^{+} + \mathrm{H}) - H_{22}(\mathrm{X} + \mathrm{H}^{+}) - \Delta \mathrm{E}.
\label{deltaH}
\end{equation}
For the potential $H_{11}(\mathrm{X} + \mathrm{H}^{+})$, the polarizability term (see Equation~\ref{h11}) should include a quadrupole term in addition to the dipole polarizability if X is not an $S$ ground term \citep{gentry77}:
\begin{equation}
\frac{\alpha_d}{2R^4} \rightarrow \frac{\alpha_d}{2R^4} \pm \frac{Q}{2R^3}.
\end{equation}
To determine $\Delta H(R)$ for each element, static average electric dipole polarizabilities were taken from \citet{lide02} and quadrupole matrix elements from Tables~2 and 3 of \citet{froese77}.

The transition probability for the reaction is $p=e^{-w}$ with \citep{demkov64, swartz94}
\begin{equation}
w = \left[ 1 + \mathrm{exp}\left( \frac{2\pi^2\Delta U(R_{\mathrm{x}})}{h\alpha v} \right) \right]^{-1},
\label{demkovw}
\end{equation}
where $h$ is Planck's constant and $\alpha$ is the polarizability of the neutral species X.  Defining
\begin{equation}
\delta = \frac{\pi^2\Delta U(R_{\mathrm{x}})}{h\alpha v},
\end{equation}
the cross section at energy $E$ is 
\begin{equation}
\sigma(E) = \pi R_{\mathrm{x}}^2 \int_{1}^{\infty} \frac{1}{x^3} \left( \frac{4e^{-\delta x}}{(1 + e^{\delta x})^2} \right)dx.
\label{demkov_sigma}
\end{equation}

We compute two-channel cross sections, which can be added to determine the total cross section.  While this addition can lead to overestimated total cross sections (see Sect.~\ref{lz}), only Se and Br have more than one exoergic final channel, and their cross sections are typically small in the relevant temperature range.  As for the MCLZ calculations, the cross sections are integrated over a Maxwellian velocity distribution to derive rate coefficients.

\section{Results}\label{results}

\addtocounter{table}{2}

In Table~\ref{channels},\footnote{Table~\ref{channels} is available in the online version of this article.} we provide information for the exit channels of our MCLZ and Demkov calculations, including energy defects $\Delta$E, avoided crossing distances (or interaction distances, for singly-charged ions) $R_{\rm x}$, and the energy separation $\Delta U$ of the adiabatic curves at $R_{\rm x}$.  These parameters were used to compute the CT cross section for each final channel.  In addition, we list the molecular symmetries common to the initial and final channels and whether the reaction is Type~I or Type~II.  When a final state can connect to the initial state via more than one molecular symmetry, the cross sections for each symmetry were added to determine the state-specific cross section.  The state-resolved cross sections were added to determine the total CT cross section, and rate coefficients determined by integrating over a Maxwellian velocity distribution.

\begin{figure}
  \resizebox{\hsize}{!}{\includegraphics{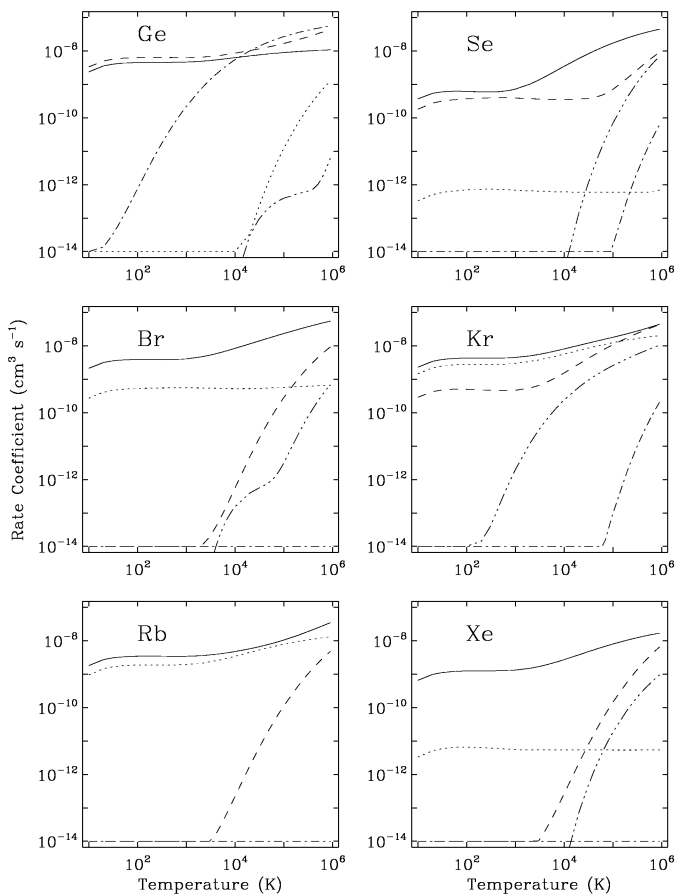}}
  \caption{Total rate coefficients for CT recombination of X$^+$ (dash-dot-dot-dot-dash curves), X$^{2+}$ (dash-dot-dash curves), X$^{3+}$ (dotted curves), X$^{4+}$ (solid curves), and X$^{5+}$ (dashed curves), where X is the element indicated in each panel.  The $q=5$ rate coefficients are lower limits due to incomplete experimental energy level measurements.  Note that the Rb$^+$ rate coefficient is below 10$^{-14}$~cm$^3$\,s$^{-1}$ over the plotted temperature range.} \label{ctr_rates}
\end{figure}

The total CT recombination rate coefficients of each ion are given in Table~\ref{total_recomb} as functions of temperature, and are illustrated in Figure~\ref{ctr_rates}.  We note that for $q=5$, highly-excited Rydberg states can be favorable exit channels, but the reported level energies in NIST for X$^{4+}$ ions are incomplete at those energies.  Thus, CT into those highly-excited states could not be computed, and as a result the rate coefficients for $q=5$ are lower limits.  As seen in Fig.~\ref{ctr_rates}, the computed $q=5$ rate coefficients are smaller than those for $q=4$ with the exception of Ge$^{5+}$.  Since CT rate coefficients generally increase with charge for $q\geq 3$, we recommend that $q=4$ CT rate coefficients be used as rough estimates for Se, Br, Kr, Rb, and Xe $q\geq 5$ ions.

When the radial coupling CT rate coefficient from our MCLZ and Demkov calculations is sufficiently small, CT can proceed via other coupling mechanisms.  Two such examples are radiative CT and spin-orbit (SO) coupling.  For radiative CT, the canonical rate coefficient is taken to be 10$^{-14}$~cm$^3$~s$^{-1}$ \citep{butler77}.  This value is uncertain by at least an order of magnitude \citep{butler77, stancil96}, but given the small rate coefficient, CT is unlikely to be an important recombination mechanism compared to radiative or dielectronic recombination when radiative CT dominates radial coupling.  The rate coefficients listed in the tables are the maximum of the radially-coupled and radiative CT rate coefficients (thus, all values of 1.00E-14 indicate that radiative CT dominates radially-coupled CT).  For some exit channels, radiative CT dominates radial coupling at all temperatures from 10$^2$--10$^6$~K.  States with such small rate coefficients tend to be associated with small or large avoided crossing distances ($R_{\rm x} \lesssim 5$, $R_{\rm x} \gtrsim 15$--20~$a_0$).  For some singly- and doubly-charged ions, the only exothermic exit channel(s) has a very small rate coefficient dominated by radiative CT.

Spin-orbit coupling is not as well understood as the radial or radiative mechanisms.  Rate coefficients for SO-coupled CT reactions can range from 10$^{-17}$--10$^{-12}$~cm$^3$~s$^{-1}$ \citep{bd79, bd80b, pradhan94, stancil01}, though detailed calculations are very few and to our knowledge exist only for singly-charged ions.  Because of this, we consider SO coupling only for CT recombination of singly-charged ions and CT ionization of neutral species.  For small radial coupling CT rate coefficients ($\leq 10^{-12}$cm$^3$~s$^{-1}$), we adopt 10$^{-12}$~cm$^3$~s$^{-1}$, which is the largest computed rate coefficient for SO coupling \citep{pradhan94}, as an upper limit to the rate coefficient.  This estimate provides a much smaller, but more robust upper limit than estimates based on the Langevin model \citep{stancil01}.

However, SO coupling is not relevant for all of the singly-ionized species we consider, since it only connects initial and final states of different molecular symmetries for which radial and rotational coupling are forbidden.  For CT recombination of Rb$^+$ (and the reverse reaction), the entrance and exit channels correspond to $^2\Sigma^+$ molecular states, and SO coupling therefore plays no role.  In the case of Kr$^+$ and Xe$^+$, for which there is only one exothermic exit channel, the diabatic potentials for the triplet exit channels do not cross the potential curve of the singlet entrance channel, and hence SO coupling is unlikely to be important for these reactions.  We therefore adopt the SO coupling upper limit to the rate coefficients only for singly-ionized Ge, Se, and Br.

\begin{figure}
  \resizebox{\hsize}{!}{\includegraphics{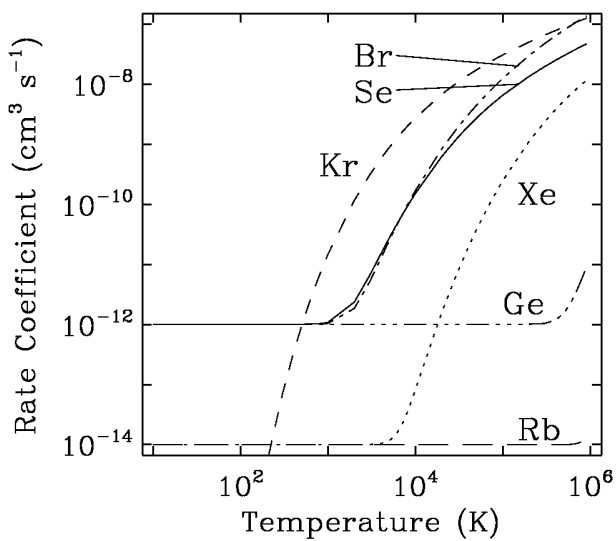}}
  \caption{Total rate coefficients for CT ionization of the indicated neutral species.} \label{cti_rates}
\end{figure}

CT ionization is exothermic for the neutrals of the elements we consider, with the exception of Kr.  We provide rate coefficients for these systems in Table~\ref{total_ion}, and depict them in Figure~\ref{cti_rates}.  As for CT recombination, we adopt the largest of the rate coefficients associated with the radial, radiative, and (when important) SO mechanisms.

\begin{table*}
\centering
\caption{Total charge transfer ionization rate coefficients (cm$^3$~s$^{-1}$) for the reactions X$^0$~+~H$^+$~$\rightarrow$~X$^{+}$~+~H~+~$\delta E$.}\label{total_ion}
\begin{tabular}{lcccccccccc}
\hline \hline
 & \multicolumn{10}{c}{$T$~(K)} \\ \cline{2-11}
 & 1.0E+02 & 2.0E+02 & 4.0E+02 & 6.0E+02 & 8.0E+02 & 1.0E+03 & 2.0E+03 & 4.0E+03 & 6.0E+03 & 8.0E+03 \\
Projectile & 1.0E+04 & 2.0E+04 & 4.0E+04 & 6.0E+04 & 8.0E+04 & 1.0E+05 & 2.0E+05 & 4.0E+05 & 6.0E+05 & 8.0E+05 \\
\hline
Ge$^0$ & 1.01E-12 & 1.01E-12 & 1.01E-12 & 1.01E-12 & 1.01E-12 & 1.01E-12 & 1.01E-12 & 1.01E-12 & 1.01E-12 & 1.01E-12 \\
 & 1.01E-12 & 1.01E-12 & 1.01E-12 & 1.01E-12 & 1.01E-12 & 1.01E-12 & 1.01E-12 & 1.18E-12 & 2.32E-12 & 5.83E-12 \\
Se$^0$ & 1.01E-12 & 1.01E-12 & 1.01E-12 & 1.01E-12 & 1.01E-12 & 1.01E-12 & 1.01E-12 & 1.01E-12 & 1.01E-12 & 1.02E-12 \\
 & 1.04E-12 & 1.86E-12 & 1.24E-11 & 4.24E-11 & 9.51E-11 & 1.71E-10 & 8.41E-10 & 3.11E-09 & 6.00E-09 & 9.19E-09 \\
Br$^0$ & 1.01E-12 & 1.01E-12 & 1.01E-12 & 1.02E-12 & 1.04E-12 & 1.09E-12 & 2.37E-12 & 1.48E-11 & 4.42E-11 & 9.07E-11 \\
 & 1.52E-10 & 6.28E-10 & 1.98E-09 & 3.51E-09 & 5.05E-09 & 6.58E-09 & 1.36E-08 & 2.50E-08 & 3.45E-08 & 4.29E-08 \\
Kr$^0$ & 3.62E-45 & 1.80E-31 & 1.42E-23 & 1.62E-20 & 8.27E-19 & 1.09E-17 & 4.93E-15 & 3.27E-13 & 2.14E-12 & 6.70E-12 \\
 & 1.48E-11 & 1.17E-10 & 5.70E-10 & 1.22E-09 & 1.96E-09 & 2.75E-09 & 6.81E-09 & 1.41E-08 & 2.04E-08 & 2.59E-08 \\
Rb$^0$ & 1.00E-14 & 1.00E-14 & 1.00E-14 & 1.00E-14 & 1.00E-14 & 1.00E-14 & 1.00E-14 & 1.00E-14 & 1.00E-14 & 1.00E-14 \\
 & 1.00E-14 & 1.00E-14 & 1.00E-14 & 1.00E-14 & 1.00E-14 & 1.00E-14 & 1.00E-14 & 1.00E-14 & 1.01E-14 & 1.12E-14 \\
Xe$^0$ & 1.00E-14 & 1.00E-14 & 1.00E-14 & 1.00E-14 & 1.00E-14 & 1.00E-14 & 1.00E-14 & 1.04E-14 & 1.48E-14 & 3.34E-14 \\
 & 8.32E-14 & 1.60E-12 & 1.91E-11 & 6.51E-11 & 1.42E-10 & 2.49E-10 & 1.13E-09 & 3.78E-09 & 6.84E-09 & 9.99E-09 \\
\hline
\end{tabular}
\tablefoot{For Ge, Se, and Br, the spin-orbit coupling rate coefficient found for Cl$^+$, 10$^{-12}$~cm$^3$~s$^{-1}$ \citep{pradhan94}, is adopted as an upper limit when the rate from radial coupling falls below that value.  SO coupling is not relevant for CT reactions involving neutral Kr, Rb, and Xe, and hence we set the rate coefficient to the canonical radiative CT rate \citep[10$^{-14}$~cm$^3$~s$^{-1}$;][]{butler77} when the radial coupling rate coefficient is smaller than that value.   Note that radiative CT does not occur for the endoergic reaction Kr$^0$~+~H$^+$~$\rightarrow$~Kr$^{+}$~+~H~+~$\delta E$.}
\end{table*}

The CT recombination rate coefficients for singly-ionized species were computed from the CT ionization rate coefficients via detailed balance:
\begin{equation}
\alpha_{\rm rec} = \alpha_{\rm ion} \frac{p_{0, \rm rec}}{p_{0, \rm ion}} \mathrm{exp}\left( \frac{\Delta E_{\infty}}{k_{\rm B}T} \right) ,
\label{reverse}
\end{equation}
where $\alpha_{\rm rec, ion}$ are the rate coefficients for CT recombination and ionization, respectively, $p_{0, \rm rec, ion}$ are the approach probabilities corresponding to the molecular symmetries involved in the recombination and ionization reactions, $\Delta E_{\infty}$ is the asymptotic energy defect, and $k_{\rm B}$ the Boltzmann constant.  We note that Equation~\ref{reverse} was applied only to state-resolved rate coefficients.  In the case of Kr, for which CT recombination of singly-charged ions is exothermic, the CT ionization rate coefficients were computed in an analogous manner.  For neutral Kr, the energy defect is very small (corresponding to 4\,700~K), and hence the CT ionization reaction can operate efficiently in photoionized nebulae.  CT recombination of singly-charged ions is less important at photoionized temperatures for the other elements (see Fig.~\ref{ctr_rates}), as the lowest energy defect is that of Xe (corresponding to 17\,000~K).

State-resolved rate coefficients are important for understanding the emission line spectra of photoionized nebulae, since CT populates excited states from which emission lines can be measured \citep[e.g.,][]{kf96}.  We therefore provide final-state resolved rate coefficients for CT recombination in Table~4 and for CT ionization in Table~5 (both available at the CDS).

The derived abundances of Se and Kr have explicitly been shown to be sensitive to uncertainties in CT rate coefficients \citep{sterling07}, and other trans-iron elements are expected to be similarly sensitive.  Unfortunately, it is not possible to ascertain robust uncertainties for CT rate coefficients computed with the LZ or Demkov approximations without accompanying QMOCC calculations.  \citet{bd80} found that their LZ rate coefficients agreed with quantal calculations to within a factor of three for ions of light elements with large CT rate coefficients, but that the disagreement could exceed an order of magnitude in the case of systems with small rate coefficients.  No such comparison is possible for the heavy ions we have investigated, since QMOCC computations are not available for any of these systems.  Our forthcoming study of the sensitivity of \emph{n}-capture abundance determinations to atomic data uncertainties, using Monte Carlo simulations with photoionization modeling codes, will help to identify the ions that most urgently require QMOCC determinations of CT rate coefficients.

\section{Summary}\label{summ}

We have presented CT recombination rate coefficients for the first five ionization stages of the \emph{n}-capture elements Ge, Se, Br, Kr, Rb, and Xe reacting with neutral hydrogen.  A multi-channel Landau-Zener approach was employed for multiply-charged ions, while the Demkov approximation was used for singly-charged species.  We note that the computed rate coefficients for $q=5$ are lower limits, due to the incompleteness of energy listings in the NIST database.  Because of that, we recommend that the four-times ionized recombination rate coefficients be used for $q\geq 5$, with the exception of Ge$^{5+}$.  For reactions between the neutral species of these elements and H$^+$, we computed CT ionization rate coefficients.  Total rate coefficients are given over the range of temperatures 10$^2$--10$^6$~K in Tables~\ref{total_recomb} and \ref{total_ion}, while final-state resolved rate coefficients are presented in Tables~4 and 5 (available at the CDS).  In addition, all rate coefficients can be obtained at http://www.pa.msu.edu/astro/atomicdata/charge\_transfer/.  In tandem with our concurrent investigations into the photoionization and radiative and dielectronic recombination properties of \emph{n}-capture elements \citep[e.g.,][]{sterling11b, sterling11c}, these CT rate coefficient determinations will enable the abundances of \emph{n}-capture elements to be reliably determined in ionized astrophysical nebulae.  Such studies bear significant implications for \emph{n}-capture nucleosynthesis, the structure and internal mixing of evolved stars, and the chemical evolution of trans-iron elements in the Universe.

\acknowledgements

N.\ C.\ Sterling gratefully acknowledges support from an NSF Astronomy and Astrophysics Postdoctoral Fellowship under award AST-0901432 and from NASA grant 06-APRA206-0049.  The work of P.\ C.\ Stancil was partially supported by NASA grant NNX09AC46G.

\bibliographystyle{aa}

\bibliography{sterling_ct.bib}

\clearpage 

\renewcommand{\thefootnote}{\alph{footnote}}
\onllongtab{1}{
\centering
\begin{longtable}{llcccc}
\caption{Details of MCLZ and Demkov calculations, including exit channels, molecular symmetries, energy defects $\Delta E$, avoided crossing or interaction distances $R_{\rm x}$, and the energy separations of the adiabatic potentials $\Delta U(R_{\rm x})$.  The molecular symmetries of each entrance channel is given in parentheses after the reaction.}\label{channels} \\
\hline \hline
Exit & $\Delta E$ & $R_{\mathrm{x}}$ & $\Delta U(R_{\rm x})$ & Molecular &  \\
Channels\footnotemark[1] & (a.u.) & (a.u.) & (a.u.) & Symmetries\footnotemark[2] & Type \\
\hline
\multicolumn{6}{c}{Ge$^0$($^3$P) + H$^+$($^1$S) $\rightleftharpoons$ Ge$^+$ + H($^2$S) + $\Delta E$ \,\,\,\, ($^3\Sigma^-$, $^3\Pi$)} \\
\hline
\endfirsthead
\caption{Continued.} \\
\hline
\endhead
\hline
\endfoot
\hline
\endlastfoot
4s$^2$\,4p $^2$P$^o$ & 0.2041 & 7.258 & 0.2089 & $^3\Sigma^-$ & I \\
4s$^2$\,4p $^2$P$^o$  & 0.2041 & 7.406 & 0.1943 & $^3\Pi$ & I \\
\hline
\multicolumn{6}{c}{Ge$^{2+}$($^1$S) + H($^2$S) $\rightarrow$ Ge$^+$ + H$^+$($^1$S) + $\Delta E$ \,\,\,\, ($^2\Sigma^+$)} \\
\hline
4s$^2$\,4p $^2$P$^o$  & 0.0805 & 12.48 & 5.922E-03 & $^2\Sigma^+$ & I \\
\hline
\multicolumn{6}{c}{Ge$^{3+}$($^2$S) + H($^2$S) $\rightarrow$ Ge$^{2+}$ + H$^+$($^1$S) + $\Delta E$ \,\,\,\, ($^1\Sigma^+$, $^3\Sigma^+$)} \\
\hline
4s\,4p $^3$P$^o$  & 0.4633 & 4.648 & 0.2070 & $^3\Sigma^+$ & I \\
4s\,4p $^1$P$^o$  & 0.3333 & 6.240 & 7.592E-02 & $^1\Sigma^+$ & I \\
\hline
\multicolumn{6}{c}{Ge$^{4+}$($^1$S) + H($^2$S) $\rightarrow$ Ge$^{3+}$ + H$^+$($^1$S) + $\Delta E$ \,\,\,\, ($^2\Sigma^+$)} \\
\hline
4p $^2$P$^o$  & 0.8013 & 4.138 & 0.2732 & $^2\Sigma^+$ & I \\
4d $^2$D & 0.3111 & 9.768 & 5.463E-03 & $^2\Sigma^+$ & I \\
5s $^2$S & 0.2724 & 11.11 & 1.847E-03 & $^2\Sigma^+$ & I \\
\hline
\multicolumn{6}{c}{Ge$^{5+}$($^2$D) + H($^2$S) $\rightarrow$ Ge$^{4+}$ + H$^+$($^1$S) + $\Delta E$ \,\,\,\, ($^1\Sigma^+$, $^1\Pi$, $^1\Delta$, $^3\Sigma^+$, $^3\Pi$, $^3\Delta$)} \\
\hline
3d$^9$\,4d $^3$S & 0.7481 & 5.723 & 0.1071 & $^3\Sigma^+$ & I \\
3d$^9$\,4d $^1$P & 0.7219 & 5.901 & 9.530E-02 & $^1\Pi$ & I \\
3d$^9$\,4d $^3$G & 0.7202 & 5.913 & 9.454E-02 & $^3\Sigma^+$, $^3\Pi$, $^3\Delta$ & I \\
3d$^9$\,4d $^3$P & 0.7161 & 5.942 & 9.274E-02 & $^3\Pi$ & I \\
3d$^9$\,4d $^3$D & 0.7095 & 5.990 & 8.983E-02 & $^3\Sigma^+$, $^3\Pi$, $^3\Delta$ & I \\
3d$^9$\,4d $^1$F & 0.6993 & 6.065 & 8.544E-02 & $^1\Pi$, $^1\Delta$ & I \\
3d$^9$\,4d $^1$D & 0.6905 & 6.132 & 8.168E-02 & $^1\Sigma^+$, $^1\Pi$, $^1\Delta$ & I \\
3d$^9$\,4d $^3$F & 0.6903 & 6.134 & 8.157E-02 & $^3\Pi$, $^3\Delta$ & I \\
3d$^9$\,5s $^3$D & 0.5888 & 7.062 & 4.274E-02 & $^3\Sigma^+$, $^3\Pi$, $^3\Delta$ & I \\
3d$^9$\,4d $^1$S & 0.5763 & 7.200 & 3.870E-02 & $^1\Sigma^+$ & I \\
3d$^9$\,5s $^3$P & 0.5677 & 7.299 & 3.603E-02 & $^3\Pi$ & I \\
3d$^9$\,5p $^3$P$^o$  & 0.4370 & 9.313 & 7.827E-03 & $^3\Sigma^+$, $^3\Pi$ & I \\
3d$^9$\,5p $^3$F$^o$  & 0.4313 & 9.430 & 7.139E-03 & $^3\Sigma^+$, $^3\Pi$, $^3\Delta$ & I \\
3d$^9$\,5p $^1$D$^o$  & 0.4277 & 9.506 & 6.724E-03 & $^1\Pi$, $^1\Delta$ & I \\
3d$^9$\,5p $^3$D$^o$  & 0.4149 & 9.786 & 5.385E-03 & $^3\Pi$, $^3\Delta$ & I \\
3d$^9$\,5p $^1$P$^o$  & 0.4136 & 9.815 & 5.262E-03 & $^1\Sigma^+$, $^1\Pi$ & I \\
3d$^9$\,5p $^1$F$^o$  & 0.4107 & 9.882 & 4.989E-03 & $^1\Sigma^+$, $^1\Pi$, $^1\Delta$ & I \\
\hline
Exit & $\Delta E$ & $R_{\mathrm{x}}$ & $\Delta U(R_{\rm x})$ & Molecular &  \\
Channels\footnotemark[1] & (a.u.) & (a.u.) & (a.u.) & Symmetries\footnotemark[2] & Type \\
\hline
\multicolumn{6}{c}{Se$^0$($^3$P) + H$^+$($^1$S) $\rightleftharpoons$ Se$^+$ + H($^2$S) + $\Delta E$ \,\,\,\, ($^3\Sigma^-$, $^3\Pi$)} \\
\hline
4s$^2$\,4p$^3$ $^4$S$^o$  & 0.1413 & 7.006 & 0.1304 & $^3\Sigma^-$ & I \\
4s$^2$\,4p$^3$ $^2$D$^o$  & 0.0796 & 7.860 & 7.958E-02 & $^3\Pi$ & I \\
4s$^2$\,4p$^3$ $^2$D$^o$  & 0.0796 & 8.020 & 7.240E-02 & $^3\Sigma^-$ & I \\
4s$^2$\,4p$^3$ $^2$P$^o$  & 0.0338 & 9.251 & 3.396E-02 & $^3\Pi$ & I \\
\hline
\multicolumn{6}{c}{Se$^{2+}$($^3$P) + H($^2$S) $\rightarrow$ Se$^+$ + H$^+$($^1$S) + $\Delta E$ \,\,\,\, ($^2\Sigma^-$, $^2\Pi$, $^4\Sigma^-$, $^4\Pi$)} \\
\hline
4s$^2$\,4p$^3$ $^2$P$^o$ & 0.1714 & 6.040 & 8.688E-02 & $^2\Pi$ & I \\
\hline
\multicolumn{6}{c}{Se$^{3+}$($^2$P$^o$) + H($^2$S) $\rightarrow$ Se$^{2+}$ + H$^+$($^1$S) + $\Delta E$ \,\,\,\, ($^1\Sigma^+$, $^1\Pi$, $^3\Sigma^+$, $^3\Pi$)} \\
\hline
4s$^2$\,4p$^2$ $^1$S & 0.5034 & 4.256 & 0.2568 & $^1\Sigma^+$ & I \\
4s\,4p$^3$ $^3$D$^o$  & 0.2037 & 9.921 & 4.588E-05 & $^3\Pi$ & II \\
\hline
\multicolumn{6}{c}{Se$^{4+}$($^1$S) + H($^2$S) $\rightarrow$ Se$^{3+}$ + H$^+$($^1$S) + $\Delta E$ \,\,\,\, ($^2\Sigma^+$)} \\
\hline
4s\,4p$^2$ $^2$D & 0.6023 & 5.328 & 8.177E-03 & $^2\Sigma^+$ & II \\
4s\,4p$^2$ $^2$S & 0.4917 & 6.374 & 2.706E-03 & $^2\Sigma^+$ & II \\
4s$^2$\,4d $^2$D & 0.3793 & 8.088 & 2.010E-02 & $^2\Sigma^+$ & I \\
4s$^2$\,5s $^2$S & 0.3620 & 8.452 & 1.525E-02 & $^2\Sigma^+$ & I \\
4s$^2$\,5p $^2$P$^o$ & 0.2095 & 14.38 & 1.178E-04 & $^2\Sigma^+$ & I \\
\hline
\multicolumn{6}{c}{Se$^{5+}$($^2$S) + H($^2$S) $\rightarrow$ Se$^{4+}$ + H$^+$($^1$S) + $\Delta E$ \,\,\,\, ($^1\Sigma^+$, $^3\Sigma^+$)} \\
\hline
4s\,4d $^1$D & 1.0394 & 4.342 & 0.2453 & $^1\Sigma^+$ & I \\
4s\,4d $^3$D & 0.8359 & 5.209 & 0.1483 & $^3\Sigma^+$ & I \\
4s\,4d $^3$D & 0.7012 & 6.051 & 8.625E-02 & $^3\Sigma^+$ & I \\
4s\,4f $^1$F$^o$ & 0.3639 & 11.11 & 1.854E-03 & $^1\Sigma^+$ & I \\
\hline
Exit & $\Delta E$ & $R_{\mathrm{x}}$ & $\Delta U(R_{\rm x})$ & Molecular &  \\
Channels\footnotemark[1] & (a.u.) & (a.u.) & (a.u.) & Symmetries\footnotemark[2] & Type \\
\hline
\multicolumn{6}{c}{Br$^0$($^2$P$^o$) + H$^+$($^1$S) $\rightleftharpoons$ Br$^+$ + H($^2$S) + $\Delta E$ \,\,\,\, ($^2\Sigma^+$, $^2\Pi$)} \\
\hline
4s$^2$\,4p$^4$ $^3$P & 0.0589 & 7.241 & 6.159E-02 & $^2\Pi$ & I \\
4s$^2$\,4p$^4$ $^1$D & 0.0105 & 9.612 & 1.191E-02 & $^2\Pi$ & I \\
4s$^2$\,4p$^4$ $^1$D & 0.0105 & 10.05 & 8.681E-03 & $^2\Sigma^+$ & I \\
\hline
\multicolumn{6}{c}{Br$^{2+}$($^4$S$^o$) + H($^2$S) $\rightarrow$ Br$^+$ + H$^+$($^1$S) + $\Delta E$ \,\,\,\, ($^3\Sigma^-$, $^5\Sigma^-$)} \\
\hline
4s$^2$\,4p$^4$ $^3$P & 0.2947 & 3.576 & 0.3579 & $^3\Sigma^-$ & I \\
\hline
\multicolumn{6}{c}{Br$^{3+}$($^3$P) + H($^2$S) $\rightarrow$ Br$^{2+}$ + H$^+$($^1$S) + $\Delta E$ \,\,\,\, ($^2\Sigma^-$, $^2\Pi$, $^4\Sigma^-$, $^4\Pi$)} \\
\hline
4s\,4p$^4$ $^2$D & 0.4068 & 5.219 & 9.140E-03 & $^2\Pi$ & II \\
4s$^2$\,4p$^2$\,5s $^4$P & 0.1462 & 13.73 & 2.046E-04 & $^4\Sigma^-$, $^4\Pi$ & I \\
4s$^2$\,4p$^2$\,5s $^2$P & 0.1346 & 14.91 & 7.473E-05 & $^2\Sigma^-$, $^2\Pi$ & I \\
\hline
\multicolumn{6}{c}{Br$^{4+}$($^2$P$^o$) + H($^2$S) $\rightarrow$ Br$^{3+}$ + H$^+$($^1$S) + $\Delta E$ \,\,\,\, ($^1\Sigma^+$, $^1\Pi$, $^3\Sigma^+$, $^3\Pi$)} \\
\hline
4s\,4p$^3$ $^3$D$^o$ & 0.5728 & 5.570 & 6.369E-03 & $^3\Pi$ & II \\
4s\,4p$^3$ $^3$P$^o$ & 0.5091 & 6.176 & 3.352E-03 & $^3\Sigma^+$, $^3\Pi$ & II \\
4s\,4p$^3$ $^1$D$^o$ & 0.5002 & 6.277 & 3.006E-03 & $^1\Pi$ & II \\
4s$^2$\,4p\,4d $^3$F$^o$ & 0.4212 & 7.336 & 3.507E-02 & $^3\Sigma^+$, $^3\Pi$ & I \\
4s\,4p$^3$ $^1$P$^o$ & 0.4157 & 7.426 & 8.421E-04 & $^1\Sigma^+$, $^1\Pi$ & II \\
4s$^2$\,4p\,4d $^3$D$^o$ & 0.3800 & 8.074 & 2.031E-02 & $^3\Pi$ & I \\
4s$^2$\,4p\,4d $^3$P$^o$ & 0.3468 & 8.803 & 1.165E-02 & $^3\Sigma^+$, $^3\Pi$ & I \\
4s$^2$\,4p\,4d $^1$F$^o$ & 0.3448 & 8.852 & 1.121E-02 & $^1\Sigma^+$, $^1\Pi$ & I \\
4s$^2$\,4p\,4d $^1$P$^o$ & 0.3301 & 9.227 & 8.373E-03 & $^1\Sigma^+$, $^1\Pi$ & I \\
4s$^2$\,4p\,4d $^1$D$^o$ & 0.3245 & 9.380 & 7.426E-03 & $^1\Pi$ & I \\
4s$^2$\,4p\,5s $^3$P$^o$ & 0.3180 & 9.564 & 6.422E-03 & $^3\Sigma^+$, $^3\Pi$ & I \\
4s$^2$\,4p\,5s $^1$P$^o$ & 0.3035 & 10.04 & 4.526E-03 & $^1\Sigma^+$, $^1\Pi$ & I \\
4s$^2$\,4p\,5p $^3$D & 0.1619 & 18.57 & 2.983E-06 & $^3\Sigma^+$, $^3\Pi$ & I \\
\hline
\multicolumn{6}{c}{Br$^{5+}$($^1$S) + H($^2$S) $\rightarrow$ Br$^{4+}$ + H$^+$($^1$S) + $\Delta E$ \,\,\,\, ($^2\Sigma^+$)} \\
\hline
4s\,4p$^2$ $^2$D & 1.1321 & 4.040 & 2.853E-02 & $^2\Sigma^+$ & II \\
4s$^2$\,4d $^2$D & 0.8360 & 5.209 & 0.1483 & $^2\Sigma^+$ & I \\
4s$^2$\,5s $^2$S & 0.7214 & 5.904 & 9.511E-02 & $^2\Sigma^+$ & I \\
\hline
Exit & $\Delta E$ & $R_{\mathrm{x}}$ & $\Delta U(R_{\rm x})$ & Molecular &  \\
Channels\footnotemark[1] & (a.u.) & (a.u.) & (a.u.) & Symmetries\footnotemark[2] & Type \\
\hline
\multicolumn{6}{c}{Kr$^+$($^2$P$^o$) + H($^2$S) $\rightleftharpoons$ Kr + H$^+$($^1$S) + $\Delta E$ \,\,\,\, ($^1\Sigma^+$, $^1\Pi$, $^3\Sigma^+$, $^3\Pi$)} \\
\hline
4s$^2$\,4p$^6$ $^1$S & 0.0147 & 8.394 & 1.595E-02 & $^1\Sigma^+$ & I \\
\hline
\multicolumn{6}{c}{Kr$^{2+}$($^3$P) + H($^2$S) $\rightarrow$ Kr$^+$ + H$^+$($^1$S) + $\Delta E$ \,\,\,\, ($^2\Sigma^-$, $^2\Pi$, $^4\Sigma^-$, $^4\Pi$)} \\
\hline
4s$^2$\,4p$^5$ $^2$P$^o$ & 0.3873 & 2.677 & 0.4928 & $^2\Pi$ & I \\
\hline
\multicolumn{6}{c}{Kr$^{3+}$($^4$S$^o$) + H($^2$S) $\rightarrow$ Kr$^{2+}$ + H$^+$($^1$S) + $\Delta E$ \,\,\,\ ($^3\Sigma^-$, $^5\Sigma^-$)} \\
\hline
4s$^2$\,4p$^4$ $^3$P & 0.8485 & 2.577 & 0.5047 & $^3\Sigma^-$ & I \\
4s$^2$\,4p$^3$\,4d $^5$D$^o$ & 0.2269 & 8.940 & 1.047E-02 & $^5\Sigma^-$ & I \\
4s$^2$\,4p$^3$\,5s $^5$S$^o$ & 0.1942 & 10.39 & 3.313E-03 & $^5\Sigma^-$ & I \\
4s$^2$\,4p$^3$\,4d $^3$D$^o$ & 0.1816 & 11.10 & 1.870E-03 & $^3\Sigma^-$ & I \\
4s$^2$\,4p$^3$\,5s $^3$S$^o$ & 0.1675 & 12.01 & 8.767E-04 & $^3\Sigma^-$ & I \\
4s$^2$\,4p$^3$\,4d $^3$G$^o$ & 0.1264 & 15.86 & 3.248E-05 & $^3\Sigma^-$ & I \\
4s$^2$\,4p$^3$\,5s $^3$D$^o$ & 0.1099 & 18.23 & 4.025E-06 & $^3\Sigma^-$ & I \\
\hline
\multicolumn{6}{c}{Kr$^{4+}$($^3$P) + H($^2$S) $\rightarrow$ Kr$^{3+}$ + H$^+$($^1$S) + $\Delta E$ \,\,\,\, ($^2\Sigma^-$, $^2\Pi$, $^4\Sigma^-$, $^4\Pi$)} \\
\hline
4s\,4p$^4$ $^4$P & 0.8787 & 3.834 & 3.429E-02 & $^4\Sigma^-$, $^4\Pi$ & II \\
4s$^2$\,4p$^2$\,4d $^2$P$_a$ & 0.6847 & 4.775 & 0.1924 & $^2\Sigma^-$, $^2\Pi$ & I \\
4s$^2$\,4p$^2$\,4d $^4$F & 0.6265 & 5.155 & 0.1533 & $^4\Sigma^-$, $^4\Pi$ & I \\
4s$^2$\,4p$^2$\,4d $^4$D & 0.6030 & 5.328 & 0.1378 & $^4\Pi$ & I \\
4s$^2$\,4p$^2$\,4d $^2$F$_a$ & 0.5937 & 5.400 & 0.1317 & $^2\Sigma^-$, $^2\Pi$ & I \\
4s\,4p$^4$ $^2$P & 0.5532 & 5.744 & 5.308E-03 & $^2\Sigma^-$, $^2\Pi$ & II \\
4s$^2$\,4p$^2$\,4d $^4$P & 0.5079 & 6.193 & 7.838E-02 & $^4\Sigma^-$, $^4\Pi$ & I \\
4s$^2$\,4p$^2$\,5s $^4$P & 0.4901 & 6.394 & 6.834E-02 & $^4\Sigma^-$, $^4\Pi$ & I \\
4s$^2$\,4p$^2$\,4d $^2$D$_a$ & 0.4719 & 6.615 & 5.864E-02 & $^2\Pi$ & I \\
4s$^2$\,4p$^2$\,5s $^2$P & 0.4691 & 6.650 & 5.723E-02 & $^2\Sigma^-$, $^2\Pi$ & I \\
4s$^2$\,4p$^2$\,4d $^2$D$_b$ & 0.4386 & 7.069 & 4.253E-02 & $^2\Pi$ & I \\
4s$^2$\,4p$^2$\,4d $^2$P$_b$ & 0.4257 & 7.266 & 3.690E-02 & $^2\Sigma^-$, $^2\Pi$ & I \\
4s$^2$\,4p$^2$\,4d $^2$F$_b$ & 0.4240 & 7.293 & 3.618E-02 & $^2\Sigma^-$, $^2\Pi$ & I \\
4s$^2$\,4p$^2$\,5s $^2$D & 0.4124 & 7.482 & 3.152E-02 & $^2\Pi$ & I \\
4s$^2$\,4p$^2$\,4d $^2$D$_c$ & 0.3710 & 8.259 & 1.766E-02 & $^2\Pi$ & I \\
4s$^2$\,4p$^2$\,5p $^2$S$^o$ & 0.3595 & 8.508 & 1.461E-02 & $^2\Sigma^-$ & I \\
4s$^2$\,4p$^2$\,5p $^4$D$^o$ & 0.3289 & 9.260 & 8.159E-03 & $^4\Sigma^-$, $^4\Pi$ & I \\
4s$^2$\,4p$^2$\,5p $^4$P$^o$ & 0.3203 & 9.498 & 6.766E-03 & $^4\Pi$ & I \\
4s$^2$\,4p$^2$\,5p $^4$S$^o$ & 0.3058 & 9.931 & 4.797E-03 & $^4\Sigma^-$ & I \\
4s$^2$\,4p$^2$\,5p $^2$D$_a^o$ & 0.3036 & 10.00 & 4.540E-03 & $^2\Sigma^-$, $^2\Pi$ & I \\
4s$^2$\,4p$^2$\,5p $^2$P$_a^o$ & 0.2884 & 10.51 & 3.011E-03 & $^2\Pi$ & I \\
4s$^2$\,4p$^2$\,5p $^2$F$^o$ & 0.2543 & 11.88 & 9.761E-04 & $^2\Pi$ & I \\
4s$^2$\,4p$^2$\,5p $^2$D$_b^o$ & 0.2512 & 12.03 & 8.658E-04 & $^2\Sigma^-$, $^2\Pi$ & I \\
4s$^2$\,4p$^2$\,5p $^2$P$_b^o$ & 0.2235 & 13.49 & 2.522E-04 & $^2\Pi$ & I \\
\hline
\multicolumn{6}{c}{Kr$^{5+}$($^2$P$^o$) + H($^2$S) $\rightarrow$ Kr$^{4+}$ + H$^+$($^1$S) + $\Delta E$ \,\,\,\, ($^1\Sigma^+$, $^1\Pi$, $^3\Sigma^+$, $^3\Pi$)} \\
\hline
4s$^2$\,4p\,4d $^3$P$^o$ & 0.9083 & 4.841 & 0.1851 & $^3\Sigma^+$, $^3\Pi$ & I \\
4s$^2$\,4p\,4d $^3$D$^o$ & 0.8781 & 4.998 & 0.1687 & $^3\Pi$ & I \\
4s$^2$\,4p\,4d $^1$F$^o$ & 0.8110 & 5.344 & 0.1364 & $^1\Sigma^+$, $^1\Pi$ & I \\
4s$^2$\,4p\,4d $^1$P$^o$ & 0.7946 & 5.437 & 0.1287 & $^1\Sigma^+$, $^1\Pi$ & I \\
4s$^2$\,4p\,5s $^3$P$^o$ & 0.7633 & 5.619 & 0.1146 & $^3\Sigma^+$, $^3\Pi$ & I \\
4s$^2$\,4p\,5s $^1$P$^o$ & 0.7342 & 5.816 & 0.1008 & $^1\Sigma^+$, $^1\Pi$ & I \\
4s$^2$\,4p\,5p $^1$P & 0.5858 & 7.094 & 4.177E-02 & $^1\Pi$ & I \\
4s$^2$\,4p\,5p $^3$D & 0.5761 & 7.202 & 3.865E-02 & $^3\Sigma^+$, $^3\Pi$ & I \\
4s$^2$\,4p\,5p $^3$P & 0.5554 & 7.446 & 3.237E-02 & $^3\Pi$ & I \\
4s$^2$\,4p\,5p $^3$S & 0.5395 & 7.646 & 2.794E-02 & $^3\Sigma^+$ & I \\
4s$^2$\,4p\,5p $^1$D & 0.5264 & 7.822 & 2.452E-02 & $^1\Sigma^+$, $^1\Pi$ & I \\
4s$^2$\,4p\,5p $^1$S & 0.4759 & 8.592 & 1.370E-02 & $^1\Sigma^+$ & I \\
\hline
Exit & $\Delta E$ & $R_{\mathrm{x}}$ & $\Delta U(R_{\rm x})$ & Molecular &  \\
Channels\footnotemark[1] & (a.u.) & (a.u.) & (a.u.) & Symmetries\footnotemark[2] & Type \\
\hline
\multicolumn{6}{c}{Rb$^0$($^2$S) + H$^+$($^1$S) $\rightleftharpoons$ Rb$^+$ + H($^2$S) + $\Delta E$ \,\,\,\, ($^2\Sigma^+$)} \\
\hline
4s$^2$\,4p$^6$ $^1$S & 0.3462 & 10.46 & 0.3332 & $^2\Sigma^+$ & I \\
\hline
\multicolumn{6}{c}{Rb$^{2+}$($^2$P$^o$) + H($^2$S) $\rightarrow$ Rb$^+$ + H$^+$($^1$S) + $\Delta E$ \,\,\,\, ($^1\Sigma^+$, $^1\Pi$, $^3\Sigma^+$, $^3\Pi$)} \\
\hline
4s$^2$\,4p$^6$ $^1$S & 0.5031 & 2.175 & 0.5374 & $^1\Sigma^+$ & I \\
\hline
\multicolumn{6}{c}{Rb$^{3+}$($^3$P) + H($^2$S) $\rightarrow$ Rb$^{2+}$ + H$^+$($^1$S) + $\Delta E$ \,\,\,\, ($^2\Sigma^-$, $^2\Pi$, $^4\Sigma^-$, $^4\Pi$)} \\
\hline
4s$^2$\,4p$^4$\,4d $^4$D & 0.2275 & 8.917 & 1.066E-02 & $^4\Pi$ & I \\
4s$^2$\,4p$^4$\,4d $^4$F & 0.1798 & 11.20 & 1.710E-03 & $^4\Sigma^-$, $^4\Pi$ & I \\
4s$^2$\,4p$^4$\,5s $^2$D & 0.1629 & 12.34 & 6.637E-04 & $^2\Pi$ & I \\
4s$^2$\,4p$^4$\,4d $^2$P & 0.1593 & 12.62 & 5.269E-04 & $^2\Sigma^-$, $^2\Pi$ & I \\
4s$^2$\,4p$^4$\,4d $^4$P & 0.1559 & 12.89 & 4.192E-04 & $^4\Sigma^-$, $^4\Pi$ & I \\
4s$^2$\,4p$^4$\,4d $^2$D & 0.1463 & 13.73 & 2.062E-04 & $^2\Pi$ & I \\
4s$^2$\,4p$^4$\,4d $^2$F & 0.1428 & 14.06 & 1.552E-04 & $^2\Sigma^-$, $^2\Pi$ & I \\
4s$^2$\,4p$^4$\,5s $^2$P & 0.1261 & 15.90 & 3.142E-05 & $^2\Sigma^-$, $^2\Pi$ & I \\
4s$^2$\,4p$^4$\,4d $^2$D & 0.1219 & 16.45 & 1.950E-05 & $^2\Pi$ & I \\
\hline
\multicolumn{6}{c}{Rb$^{4+}$($^4$S$^o$) + H($^2$S) $\rightarrow$ Rb$^{3+}$ + H$^+$($^1$S) + $\Delta E$ \,\,\,\, ($^3\Sigma^-$, $^5\Sigma^-$)} \\
\hline
4s$^2$\,4p$^3$\,4d $^5$D$^o$ & 0.6392 & 5.043 & 0.1642 & $^5\Sigma^-$ & I \\
4s$^2$\,4p$^3$\,4d $^3$D$_a^o$ & 0.5865 & 5.458 & 0.1270 & $^3\Sigma^-$ & I \\
4s$^2$\,4p$^3$\,4d $^3$F$^o$ & 0.5553 & 5.724 & 0.1070 & $^3\Sigma^-$ & I \\
4s$^2$\,4p$^3$\,4d $^3$G$^o$ & 0.5237 & 6.027 & 8.764E-02 & $^3\Sigma^-$ & I \\
4s$^2$\,4p$^3$\,5s $^5$S$^o$ & 0.4923 & 6.364 & 6.976E-02 & $^5\Sigma^-$ & I \\
4s$^2$\,4p$^3$\,5s $^3$S$^o$ & 0.4619 & 6.744 & 5.357E-02 & $^3\Sigma^-$ & I \\
4s$^2$\,4p$^3$\,4d $^3$D$_b^o$ & 0.4225 & 7.316 & 3.558E-02 & $^3\Sigma^-$ & I \\
4s$^2$\,4p$^3$\,5s $^3$D$^o$ & 0.3955 & 7.778 & 2.534E-02 & $^3\Sigma^-$ & I \\
4s$^2$\,4p$^3$\,4d $^3$S$^o$ & 0.3935 & 7.815 & 2.465E-02 & $^3\Sigma^-$ & I \\
4s$^2$\,4p$^3$\,4d $^3$D$_c^o$ & 0.3931 & 7.822 & 2.452E-02 & $^3\Sigma^-$ & I \\
4s$^2$\,4p$^3$\,5p $^5$P & 0.3123 & 9.732 & 5.622E-03 & $^5\Sigma^-$ & I \\
4s$^2$\,4p$^3$\,5p $^3$P$_a$ & 0.2901 & 10.45 & 3.161E-03 & $^3\Sigma^-$ & I \\
4s$^2$\,4p$^3$\,5p $^3$F & 0.2189 & 13.77 & 1.988E-04 & $^3\Sigma^-$ & I \\
4s$^2$\,4p$^3$\,5p $^3$P$_b$ & 0.1942 & 15.50 & 4.465E-05 & $^3\Sigma^-$ & I \\
\hline
\multicolumn{6}{c}{Rb$^{5+}$($^3$P) + H($^2$S) $\rightarrow$ Rb$^{4+}$ + H$^+$($^1$S) + $\Delta E$ \,\,\,\, ($^2\Sigma^-$, $^2\Pi$, $^4\Sigma^-$, $^4\Pi$)} \\
\hline
4s\,4p$^4$ $^4$P & 1.3832 & 3.388 & 4.999E-02 & $^4\Sigma^-$, $^4\Pi$ & II \\
4s\,4p$^4$ $^2$D & 1.2509 & 3.706 & 3.833E-02 & $^2\Pi$ & II \\
4s\,4p$^4$ $^2$P & 1.1465 & 3.996 & 2.969E-02 & $^2\Sigma^-$, $^2\Pi$ & II \\
4s$^2$\,4p$^2$\,5s $^4$P & 0.7897 & 5.465 & 0.1264 & $^4\Sigma^-$, $^4\Pi$ & I \\
4s$^2$\,4p$^2$\,5s $^2$P & 0.7817 & 5.513 & 0.1226 & $^2\Sigma^-$, $^2\Pi$ & I \\
4s$^2$\,4p$^2$\,5s $^2$D & 0.7110 & 5.979 & 9.049E-02 & $^2\Pi$ & I \\
\hline
Exit & $\Delta E$ & $R_{\mathrm{x}}$ & $\Delta U(R_{\rm x})$ & Molecular &  \\
Channels\footnotemark[1] & (a.u.) & (a.u.) & (a.u.) & Symmetries\footnotemark[2] & Type \\
\hline
\multicolumn{6}{c}{Xe$^0$($^1$S) + H$^+$($^1$S) $\rightleftharpoons$ Xe$^+$ + H($^2$S) + $\Delta E$ \,\,\,\, ($^1\Sigma^+$)} \\
\hline
5s$^2$\,5p$^5$ $^2$P$^o$ & 0.0380 & 7.933 & 3.515E-02 & $^1\Sigma^+$ & I \\
\hline
\multicolumn{6}{c}{Xe$^{2+}$($^3$P) + H($^2$S) $\rightarrow$ Xe$^+$ + H$^+$($^1$S) + $\Delta E$ \,\,\,\, ($^2\Sigma^-$, $^2\Pi$, $^4\Sigma^-$, $^4\Pi$)} \\
\hline
5s$^2$\,5p$^5$ $^2$P$^o$ & 0.2551 & 4.078 & 0.2817 & $^2\Sigma^-$, $^2\Pi$ & I \\
\hline
\multicolumn{6}{c}{Xe$^{3+}$($^4$S$^o$) + H($^2$S) $\rightarrow$ Xe$^{2+}$ + H$^+$($^1$S) + $\Delta E$ \,\,\,\, ($^3\Sigma^-$, $^5\Sigma^-$)} \\
\hline
5s$^2$\,5p$^3$\,5d $^5$D$^o$ & 0.1307 & 15.35 & 5.097E-05 & $^5\Sigma^-$ & I \\
\hline
\multicolumn{6}{c}{Xe$^{4+}$($^3$P) + H($^2$S) $\rightarrow$ Xe$^{3+}$ + H$^+$($^1$S) + $\Delta E$ \,\,\,\, ($^2\Sigma^-$, $^2\Pi$, $^4\Sigma^-$, $^4\Pi$)} \\
\hline
5s\,5p$^4$ $^4$P & 0.5315 & 5.947 & 4.283E-03 & $^4\Sigma^-$, $^4\Pi$ & II \\
5s\,5p$^4$ $^2$D & 0.4386 & 7.068 & 1.259E-03 & $^2\Pi$ & II \\
5s$^2$\,5p$^2$\,5d $^2$P$_a$ & 0.3920 & 7.843 & 2.414E-02 & $^2\Sigma^-$, $^2\Pi$ & I \\
5s$^2$\,5p$^2$\,5d $^4$F & 0.3608 & 8.479 & 1.494E-02 & $^4\Sigma^-$, $^4\Pi$ & I \\
5s$^2$\,5p$^2$\,5d $^2$F$_a$ & 0.3494 & 8.741 & 1.222E-02 & $^2\Sigma^-$, $^2\Pi$ & I \\
5s$^2$\,5p$^2$\,5d $^4$D & 0.3172 & 9.587 & 6.306E-03 & $^4\Pi$ & I \\
5s$^2$\,5p$^2$\,5d $^4$P & 0.2713 & 11.15 & 1.781E-03 & $^4\Sigma^-$, $^4\Pi$ & I \\
5s$^2$\,5p$^2$\,5d $^2$G & 0.2647 & 11.43 & 1.425E-03 & $^2\Pi$ & I \\
5s$^2$\,5p$^2$\,6s $^4$P & 0.2450 & 12.32 & 6.749E-04 & $^4\Sigma^-$, $^4\Pi$ & I \\
5s$^2$\,5p$^2$\,5d $^2$D$_a$ & 0.2437 & 12.39 & 6.397E-04 & $^2\Pi$ & I \\
5s$^2$\,5p$^2$\,6s $^2$P & 0.2237 & 13.48 & 2.548E-04 & $^2\Sigma^-$, $^2\Pi$ & I \\
5s$^2$\,5p$^2$\,5d $^2$F$_b$ & 0.1969 & 15.29 & 5.361E-05 & $^2\Sigma^-$, $^2\Pi$ & I \\
5s$^2$\,5p$^2$\,5d $^2$D$_b$ & 0.1935 & 15.55 & 4.253E-05 & $^2\Pi$ & I \\
5s$^2$\,5p$^2$\,5d $^2$P$_b$ & 0.1867 & 16.12 & 2.605E-05 & $^2\Sigma^-$, $^2\Pi$ & I \\
5s$^2$\,5p$^2$\,4f $^4$G$^o$ & 0.1645 & 18.27 & 3.870E-06 & $^4\Sigma^-$, $^4\Pi$ & I \\
\hline
\multicolumn{6}{c}{Xe$^{5+}$($^2$P$^o$) + H($^2$S) $\rightarrow$ Xe$^{4+}$ + H$^+$($^1$S) + $\Delta E$ \,\,\,\, ($^1\Sigma^+$, $^1\Pi$, $^3\Sigma^+$, $^3\Pi$)} \\
\hline
5s$^2$\,5p\,4f $^3$D & 0.9346 & 4.745 & 0.1958 & $^3\Sigma^+$, $^3\Pi$ & I \\
5s$^2$\,5p\,4f $^1$D & 0.8902 & 4.942 & 0.1744 & $^1\Sigma^+$, $^1\Pi$ & I \\
5s$^2$\,5p\,6p $^3$D & 0.7893 & 5.468 & 0.1262 & $^3\Sigma^+$, $^3\Pi$ & I \\
5s$^2$\,5p\,6p $^3$P & 0.7835 & 5.502 & 0.1235 & $^3\Pi$ & I \\
5s$^2$\,5p\,6p $^1$P & 0.7696 & 5.586 & 0.1170 & $^1\Pi$ & I \\
5s$^2$\,5p\,6p $^3$S & 0.7487 & 5.719 & 0.1074 & $^3\Sigma^+$ & I \\
5s$^2$\,5p\,6p $^1$D & 0.7361 & 5.803 & 0.1017 & $^1\Sigma^+$, $^1\Pi$ & I \\
5s$^2$\,5p\,6p $^1$S & 0.6947 & 6.100 & 8.346E-02 & $^1\Sigma^+$ & I \\
\footnotetext[1]{Exit channels refer to states in the resulting ion after CT with H or H$^+$.  Only information about electrons in the valence principal quantum number shell is given.  Some terms are given a lettered sub-index (e.g., $^2$D$_a$) if multiple terms with the same configuration and orbital and spin angular momenta are exoergic exit channels.  Note that exit channels with rate coefficients smaller than 10$^{-14}$~cm$^3$~s$^{-1}$ (the canonical radiative CT rate) in the temperature range 10$^2$--10$^6$~K are not listed, unless no exoergic exit channels with larger rate coefficients exist.}
\footnotetext[2]{Only molecular symmetries that couple to the entrance channel are listed.}
\end{longtable}
}

\onecolumn
\renewcommand{\thefootnote}{\alph{footnote}}
\centering
\begin{longtable}{lcccccccccc}
\caption{Total charge transfer recombination rate coefficients (cm$^3$~s$^{-1}$) for the reactions X$^{q+}$~+~H~$\rightarrow$~X$^{(q-1)+}$~+~H$^+$~+~$\delta E$.  Each listed rate coefficient is the largest of the radially-coupled CT rate coefficient from our MCLZ and Demkov calculations and the canonical radiative CT rate coefficient $10^{-14}$~cm$^3$~s$^{-1}$ \citep{butler77, stancil96}.  Note that radiative CT does not occur for endoergic reactions.}\label{total_recomb} \\
\hline \hline
\hline
 & \multicolumn{10}{c}{$T$~(K)} \\ \cline{2-11}
Projectile & 1.0E+02 & 2.0E+02 & 4.0E+02 & 6.0E+02 & 8.0E+02 & 1.0E+03 & 2.0E+03 & 4.0E+03 & 6.0E+03 & 8.0E+03 \\
Ion X$^{q+}$ & 1.0E+04 & 2.0E+04 & 4.0E+04 & 6.0E+04 & 8.0E+04 & 1.0E+05 & 2.0E+05 & 4.0E+05 & 6.0E+05 & 8.0E+05 \\
\hline
\endfirsthead
\caption{Continued.} \\
\hline
Projectile & \multicolumn{10}{c}{$T$~(K)} \\ \cline{2-11}
Ion & 1.0E+02 & 2.0E+02 & 4.0E+02 & 6.0E+02 & 8.0E+02 & 1.0E+03 & 2.0E+03 & 4.0E+03 & 6.0E+03 & 8.0E+03 \\
X$^{q+}$ & 1.0E+04 & 2.0E+04 & 4.0E+04 & 6.0E+04 & 8.0E+04 & 1.0E+05 & 2.0E+05 & 4.0E+05 & 6.0E+05 & 8.0E+05 \\
\hline
\endhead
\hline
\endfoot
\hline
\endlastfoot
Ge$^+$ & 0.00E+00 & 0.00E+00 & 7.94E-83 & 1.68E-59 & 7.72E-48 & 7.67E-41 & 7.59E-27 & 7.54E-20 & 1.62E-17 & 2.38E-16 \\
 & 1.19E-15 & 2.99E-14 & 1.50E-13 & 2.56E-13 & 3.35E-13 & 3.94E-13 & 5.45E-13 & 7.47E-13 & 1.56E-12 & 4.03E-12 \\
Ge$^{2+}$ & 7.67E-13 & 5.37E-12 & 3.12E-11 & 7.81E-11 & 1.42E-10 & 2.20E-10 & 7.32E-10 & 1.96E-09 & 3.21E-09 & 4.39E-09 \\
 & 5.50E-09 & 1.01E-08 & 1.64E-08 & 2.08E-08 & 2.41E-08 & 2.69E-08 & 3.58E-08 & 4.50E-08 & 5.02E-08 & 5.38E-08 \\
Ge$^{3+}$ & 1.00E-14 & 1.00E-14 & 1.00E-14 & 1.00E-14 & 1.00E-14 & 1.00E-14 & 1.00E-14 & 1.00E-14 & 1.00E-14 & 1.00E-14 \\
 & 1.00E-14 & 3.36E-14 & 6.03E-13 & 2.54E-12 & 6.37E-12 & 1.23E-11 & 7.46E-11 & 3.23E-10 & 6.66E-10 & 1.06E-09 \\
Ge$^{4+}$ & 4.50E-09 & 4.56E-09 & 4.56E-09 & 4.57E-09 & 4.60E-09 & 4.64E-09 & 4.90E-09 & 5.38E-09 & 5.76E-09 & 6.05E-09 \\
 & 6.30E-09 & 7.11E-09 & 7.93E-09 & 8.40E-09 & 8.72E-09 & 8.95E-09 & 9.63E-09 & 1.02E-08 & 1.05E-08 & 1.07E-08 \\
Ge$^{5+}$ & 6.34E-09 & 6.40E-09 & 6.37E-09 & 6.33E-09 & 6.31E-09 & 6.32E-09 & 6.52E-09 & 7.10E-09 & 7.66E-09 & 8.15E-09 \\
 & 8.58E-09 & 1.02E-08 & 1.24E-08 & 1.41E-08 & 1.55E-08 & 1.67E-08 & 2.19E-08 & 2.98E-08 & 3.61E-08 & 4.15E-08 \\
\hline
Se$^+$ & 0.00E+00 & 0.00E+00 & 0.00E+00 & 5.88E-83 & 1.24E-71 & 2.96E-64 & 4.67E-47 & 8.96E-36 & 5.41E-31 & 3.36E-28 \\
 & 2.58E-26 & 1.12E-21 & 2.28E-18 & 7.39E-17 & 6.27E-16 & 2.80E-15 & 1.43E-13 & 3.09E-12 & 1.38E-11 & 3.55E-11 \\
Se$^{2+}$ & 1.00E-14 & 1.00E-14 & 1.00E-14 & 1.00E-14 & 1.00E-14 & 1.00E-14 & 1.00E-14 & 1.00E-14 & 1.00E-14 & 1.00E-14 \\
 & 1.00E-14 & 1.00E-14 & 1.00E-14 & 1.00E-14 & 1.00E-14 & 1.55E-14 & 4.00E-13 & 5.45E-12 & 1.98E-11 & 4.50E-11 \\
Se$^{3+}$ & 6.92E-13 & 7.29E-13 & 7.31E-13 & 7.14E-13 & 6.98E-13 & 6.85E-13 & 6.55E-13 & 6.34E-13 & 6.19E-13 & 6.10E-13 \\
 & 6.05E-13 & 5.96E-13 & 5.92E-13 & 5.90E-13 & 5.90E-13 & 5.90E-13 & 5.91E-13 & 5.93E-13 & 6.06E-13 & 6.51E-13 \\
Se$^{4+}$ & 6.15E-10 & 5.96E-10 & 5.98E-10 & 6.29E-10 & 6.72E-10 & 7.23E-10 & 1.03E-09 & 1.68E-09 & 2.31E-09 & 2.91E-09 \\
 & 3.48E-09 & 5.96E-09 & 9.77E-09 & 1.27E-08 & 1.52E-08 & 1.72E-08 & 2.47E-08 & 3.35E-08 & 3.90E-08 & 4.30E-08 \\
Se$^{5+}$ & 3.70E-10 & 3.90E-10 & 4.00E-10 & 3.98E-10 & 3.94E-10 & 3.89E-10 & 3.73E-10 & 3.59E-10 & 3.54E-10 & 3.53E-10 \\
 & 3.53E-10 & 3.61E-10 & 4.04E-10 & 4.79E-10 & 5.82E-10 & 7.09E-10 & 1.56E-09 & 3.66E-09 & 5.87E-09 & 8.03E-09 \\
\hline
Br$^+$ & 1.68E-93 & 4.10E-53 & 6.40E-33 & 3.45E-26 & 8.00E-23 & 8.37E-21 & 9.15E-17 & 9.56E-15 & 4.51E-14 & 9.78E-14 \\
 & 1.56E-13 & 3.95E-13 & 6.73E-13 & 1.04E-12 & 1.81E-12 & 3.20E-12 & 2.36E-11 & 1.37E-10 & 3.27E-10 & 5.67E-10 \\
Br$^{2+}$ & 1.00E-14 & 1.00E-14 & 1.00E-14 & 1.00E-14 & 1.00E-14 & 1.00E-14 & 1.00E-14 & 1.00E-14 & 1.00E-14 & 1.00E-14 \\
 & 1.00E-14 & 1.00E-14 & 1.00E-14 & 1.00E-14 & 1.00E-14 & 1.00E-14 & 1.00E-14 & 1.00E-14 & 1.00E-14 & 1.00E-14 \\
Br$^{3+}$ & 5.30E-10 & 5.44E-10 & 5.52E-10 & 5.54E-10 & 5.55E-10 & 5.55E-10 & 5.50E-10 & 5.37E-10 & 5.29E-10 & 5.25E-10 \\
 & 5.23E-10 & 5.26E-10 & 5.41E-10 & 5.55E-10 & 5.66E-10 & 5.75E-10 & 6.04E-10 & 6.32E-10 & 6.48E-10 & 6.58E-10 \\
Br$^{4+}$ & 3.89E-09 & 3.88E-09 & 3.86E-09 & 3.90E-09 & 3.98E-09 & 4.08E-09 & 4.66E-09 & 5.79E-09 & 6.78E-09 & 7.64E-09 \\
 & 8.42E-09 & 1.15E-08 & 1.57E-08 & 1.87E-08 & 2.12E-08 & 2.34E-08 & 3.12E-08 & 4.10E-08 & 4.77E-08 & 5.28E-08 \\
Br$^{5+}$ & 1.00E-14 & 1.00E-14 & 1.00E-14 & 1.00E-14 & 1.00E-14 & 1.00E-14 & 1.00E-14 & 3.81E-14 & 1.18E-13 & 2.84E-13 \\
 & 5.68E-13 & 4.76E-12 & 3.30E-11 & 9.08E-11 & 1.76E-10 & 2.86E-10 & 1.08E-09 & 3.27E-09 & 5.68E-09 & 8.12E-09 \\
\hline
Kr$^+$ & 1.00E-14 & 1.00E-14 & 1.00E-14 & 1.00E-14 & 1.00E-14 & 1.01E-14 & 1.44E-14 & 9.96E-14 & 4.04E-13 & 1.02E-12 \\
 & 2.00E-12 & 1.24E-11 & 5.35E-11 & 1.10E-10 & 1.74E-10 & 2.41E-10 & 5.81E-10 & 1.19E-09 & 1.71E-09 & 2.17E-09 \\
Kr$^{2+}$ & 1.00E-14 & 1.00E-14 & 1.00E-14 & 1.00E-14 & 1.00E-14 & 1.00E-14 & 1.00E-14 & 1.00E-14 & 1.00E-14 & 1.00E-14 \\
 & 1.00E-14 & 1.00E-14 & 1.00E-14 & 1.00E-14 & 1.00E-14 & 1.00E-14 & 1.00E-14 & 1.00E-14 & 1.00E-14 & 1.00E-14 \\
Kr$^{3+}$ & 2.75E-09 & 2.76E-09 & 2.77E-09 & 2.84E-09 & 2.92E-09 & 3.00E-09 & 3.42E-09 & 4.14E-09 & 4.72E-09 & 5.22E-09 \\
 & 5.66E-09 & 7.30E-09 & 9.34E-09 & 1.07E-08 & 1.17E-08 & 1.25E-08 & 1.50E-08 & 1.75E-08 & 1.89E-08 & 1.98E-08 \\
Kr$^{4+}$ & 4.28E-09 & 4.30E-09 & 4.29E-09 & 4.33E-09 & 4.41E-09 & 4.50E-09 & 5.02E-09 & 5.99E-09 & 6.80E-09 & 7.49E-09 \\
 & 8.09E-09 & 1.03E-08 & 1.32E-08 & 1.51E-08 & 1.67E-08 & 1.81E-08 & 2.32E-08 & 3.06E-08 & 3.66E-08 & 4.18E-08 \\
Kr$^{5+}$ & 5.00E-10 & 4.83E-10 & 4.63E-10 & 4.58E-10 & 4.61E-10 & 4.69E-10 & 5.48E-10 & 7.72E-10 & 1.03E-09 & 1.29E-09 \\
 & 1.55E-09 & 2.86E-09 & 5.22E-09 & 7.29E-09 & 9.15E-09 & 1.08E-08 & 1.76E-08 & 2.73E-08 & 3.44E-08 & 4.03E-08 \\
\hline
Rb$^+$ & 0.00E+00 & 0.00E+00 & 0.00E+00 & 0.00E+00 & 0.00E+00 & 0.00E+00 & 0.00E+00 & 1.42E-92 & 4.23E-77 & 5.61E-68 \\
 & 8.43E-62 & 1.17E-46 & 4.32E-36 & 2.00E-31 & 1.23E-28 & 1.01E-26 & 6.58E-22 & 2.28E-18 & 1.01E-16 & 1.05E-15 \\
Rb$^{2+}$ & 1.00E-14 & 1.00E-14 & 1.00E-14 & 1.00E-14 & 1.00E-14 & 1.00E-14 & 1.00E-14 & 1.00E-14 & 1.00E-14 & 1.00E-14 \\
 & 1.00E-14 & 1.00E-14 & 1.00E-14 & 1.00E-14 & 1.00E-14 & 1.00E-14 & 1.00E-14 & 1.00E-14 & 1.00E-14 & 1.00E-14 \\
Rb$^{3+}$ & 1.88E-09 & 1.89E-09 & 1.88E-09 & 1.89E-09 & 1.91E-09 & 1.94E-09 & 2.13E-09 & 2.51E-09 & 2.85E-09 & 3.15E-09 \\
 & 3.42E-09 & 4.46E-09 & 5.79E-09 & 6.68E-09 & 7.34E-09 & 7.87E-09 & 9.56E-09 & 1.12E-08 & 1.22E-08 & 1.28E-08 \\
Rb$^{4+}$ & 3.44E-09 & 3.46E-09 & 3.42E-09 & 3.39E-09 & 3.39E-09 & 3.41E-09 & 3.56E-09 & 3.92E-09 & 4.23E-09 & 4.50E-09 \\
 & 4.75E-09 & 5.74E-09 & 7.22E-09 & 8.44E-09 & 9.51E-09 & 1.05E-08 & 1.47E-08 & 2.16E-08 & 2.73E-08 & 3.24E-08 \\
Rb$^{5+}$ & 1.00E-14 & 1.00E-14 & 1.00E-14 & 1.00E-14 & 1.00E-14 & 1.00E-14 & 1.00E-14 & 1.60E-14 & 4.82E-14 & 1.13E-13 \\
 & 2.23E-13 & 1.79E-12 & 1.24E-11 & 3.47E-11 & 6.86E-11 & 1.13E-10 & 4.61E-10 & 1.52E-09 & 2.78E-09 & 4.12E-09 \\
\hline
Xe$^+$ & 1.52E-85 & 3.53E-56 & 1.56E-39 & 3.69E-33 & 1.23E-29 & 2.41E-27 & 5.06E-22 & 1.63E-18 & 5.39E-17 & 4.36E-16 \\
 & 1.84E-15 & 7.25E-14 & 1.18E-12 & 4.44E-12 & 1.02E-11 & 1.84E-11 & 8.84E-11 & 3.06E-10 & 5.59E-10 & 8.20E-10 \\
Xe$^{2+}$ & 1.00E-14 & 1.00E-14 & 1.00E-14 & 1.00E-14 & 1.00E-14 & 1.00E-14 & 1.00E-14 & 1.00E-14 & 1.00E-14 & 1.00E-14 \\
 & 1.00E-14 & 1.00E-14 & 1.00E-14 & 1.00E-14 & 1.00E-14 & 1.00E-14 & 1.00E-14 & 1.00E-14 & 1.00E-14 & 1.00E-14 \\
Xe$^{3+}$ & 6.58E-12 & 6.35E-12 & 5.96E-12 & 5.74E-12 & 5.61E-12 & 5.53E-12 & 5.38E-12 & 5.40E-12 & 5.42E-12 & 5.43E-12 \\
 & 5.42E-12 & 5.41E-12 & 5.39E-12 & 5.37E-12 & 5.37E-12 & 5.36E-12 & 5.35E-12 & 5.39E-12 & 5.41E-12 & 5.42E-12 \\
Xe$^{4+}$ & 1.25E-09 & 1.26E-09 & 1.26E-09 & 1.28E-09 & 1.30E-09 & 1.33E-09 & 1.50E-09 & 1.85E-09 & 2.17E-09 & 2.46E-09 \\
 & 2.72E-09 & 3.80E-09 & 5.30E-09 & 6.39E-09 & 7.26E-09 & 7.99E-09 & 1.05E-08 & 1.34E-08 & 1.51E-08 & 1.64E-08 \\
Xe$^{5+}$ & 1.00E-14 & 1.00E-14 & 1.00E-14 & 1.00E-14 & 1.00E-14 & 1.00E-14 & 1.00E-14 & 2.00E-14 & 5.90E-14 & 1.37E-13 \\
 & 2.68E-13 & 2.17E-12 & 1.54E-11 & 4.39E-11 & 8.78E-11 & 1.46E-10 & 6.08E-10 & 2.03E-09 & 3.74E-09 & 5.57E-09 \\
\hline
\end{longtable}

\end{document}